\documentclass[twocolumn,apj]{emulateapj}

\usepackage{times}

\begin{document}
\shorttitle{Old Galaxies at $z\sim$1.5}
\shortauthors{McGrath, Stockton, \& Canalizo}
\journalinfo{Accepted for Publication in ApJ}
\submitted{Accepted for Publication in ApJ}

\title{Stellar Populations of Luminous Evolved Galaxies at \lowercase{$z\sim1.5$}\altaffilmark{1}}

\author{Elizabeth J. McGrath and Alan Stockton\altaffilmark{2,3}}
\affil{Institute for Astronomy, University of Hawaii, 2680 Woodlawn Dr., Honolulu, HI 96822}
\author{Gabriela Canalizo\altaffilmark{2}}
\affil{Department of Physics and Astronomy, University of California, Riverside, CA 92521}

\altaffiltext{1}{Some of the data presented herein were obtained at the W. M. Keck Observatory, which is operated as a scientific partnership among the California Institute of Technology, the University of California, and the National Aeronautics and Space Administration.  The observatory was made possible by the generous financial support of the W. M. Keck Foundation.  Results are also based in part on data collected at Subaru Telescope, which is operated by the National Astronomical Observatory of Japan, and on observations made with the 
NASA/ESA Hubble Space Telescope, obtained at the Space Telescope Science 
Institute, which is operated by the Association of Universities for Research in 
Astronomy, Inc., under NASA contract NAS 5-26555.  These observations are 
associated with program \# GO-10418.}
\altaffiltext{2}{Visiting Astronomer at the Infrared Telescope Facility, which is operated by the University of Hawaii under Cooperative Agreement no.~NCC 5-538 with the National Aeronautics and Space Administration, Science Mission Directorate, Planetary Astronomy Program.}
\altaffiltext{3}{Also at Cerro Tololo Interamerican Observatory, Casilla 603, La Serena, 
Chile.}

\begin{abstract}
Observational evidence has been mounting over the past decade that at least some luminous ($\sim2 L^{\ast}$) galaxies 
have formed nearly all of their stars within a short period of time only 1--$2\times10^9$ years 
after the Big Bang.  These are examples of the first major episodes of star formation in the Universe and provide insights into the formation of the earliest massive galaxies.  We have examined in detail the stellar populations of six $z\sim1.5$ galaxies that appear to be passively evolving, using both ground and space-based photometry covering rest-frame UV to visible wavelengths.  
In addition, we have obtained medium-resolution spectroscopy for five of the six galaxies, covering the rest-frame UV portion of the spectrum. 
Spectral synthesis modeling for four of these galaxies favors a single burst of star formation more than 1 Gyr before the observed epoch.  The other two exhibit slightly younger ages with a higher dust content and evidence for a small contribution from either recent star formation or active nuclei.  
The implied formation redshifts for the oldest of these sources are consistent with previous studies of passive galaxies at high redshift, and
improved stellar modeling has shown these results to be quite robust.  It now seems clear that any valid galaxy formation scenario must be able to account for these massive ($\sim 2 \times 10^{11} M_{\sun}$) galaxies at very early times in the Universe.
\end{abstract}


\section{Introduction}
Understanding the formation histories of the most massive galaxies has remained an important topic shaping  both observational and theoretical studies of galaxy formation and evolution.  The basic structure of a cold dark matter (CDM) cosmology requires a bottom-up formation mechanism with a late-time build-up of the most massive systems.  
Early semi-analytic models (SAMs) based on this paradigm predicted that massive galaxies did not assemble the majority of their present-day mass until $1<z<2$ (e.g., \citealp{tho99, kau00}).  Evidence for cosmic 
``downsizing'' (e.g., \citealp{cow96, jun05}) however, appears to be inconsistent with 
the 
results generated by these early hierarchical models.

More recently, 
the addition of feedback from active galactic nuclei (AGN) into SAMs
\citep{gra04,cro06,bow06,del06} 
has provided a means to shut off star formation at earlier times and allow galaxies to evolve onto the red sequence much more quickly, 
thereby producing massive ``red and dead'' galaxies at higher redshifts.
The underlying hierarchical growth of structure, however, remains a fundamental assumption throughout.  In the AGN feedback scenario proposed by \citet{cro06}, the ``quasar mode'' operates between $2<z<4$ during which time the central black hole is built up, until the more quiescent ``radio mode'' takes over, quenching star formation by a slow heating of the surrounding halo gas.

Observational constraints have largely driven the need for these corrections to SAMs and remain a key aspect for continued understanding of the processes of galaxy formation.
One method for testing predictions from these models
is to identify galaxies containing the oldest stellar populations at high redshift.  These serve as examples of the first major episodes of star formation in the universe and provide laboratories where we can study these processes in an essentially pure and unconfused environment.

\begin{figure*}[!t]
\epsscale{1.0}
\plotone{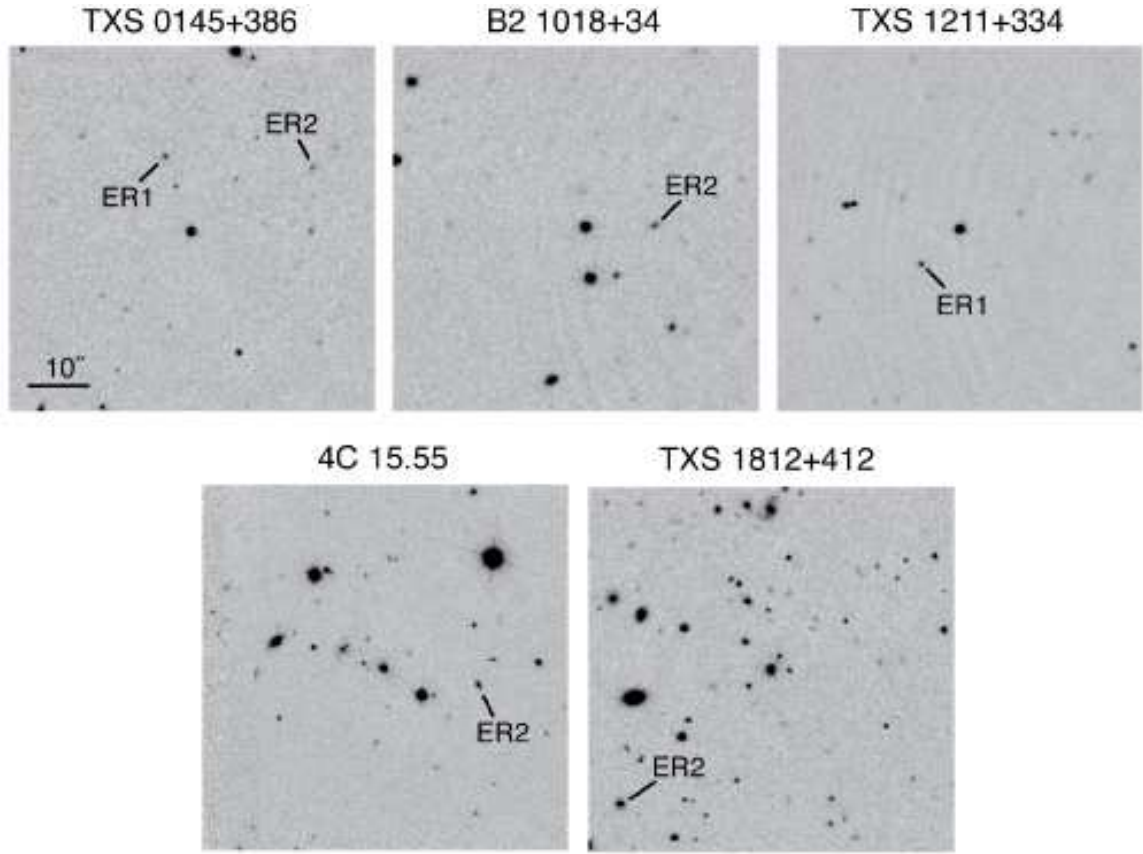}
\caption{Finding charts for the six OGs discussed in this paper.  For each field,
the quasar is centered in the image.  North is up and East is to the left.\label{fc}}
\end{figure*}

Several recent studies have discovered ``old galaxies'' (OGs; by which we refer to the age of the dominant stellar population and not necessarily the age since galaxy assembly) among extremely red objects (EROs) defined to have $R-K > 5-6$ 
\citep{liu00,dad00,dad05,cim02,cim04,iye03,yan04,fu05,sto04,sto06,sto07,stm07}.
In order to discriminate between old stellar populations and dusty starburst galaxies which contribute roughly equally to the ERO population 
\citep{cim02,yan04}, 
the strength of the 4000~{\AA} break can be used as a simple diagnostic feature.  Old stellar populations ($>1$ Gyr) show a sharp inflection at 4000~{\AA} due to metal absorption in the atmospheres of late-type stars, whereas dust-reddened starbursts show a spectral slope that is much less inclined near 4000~{\AA}.  Using the strength of this break, an approximate age can be determined for the galaxy, which can be used to extrapolate back to the formation redshift.  The results of these studies imply that a population of massive galaxies existed at very early times in the universe ($z_f > 3$) and that star formation must have proceeded extremely rapidly in these early massive galaxies.  
These conclusions are consistent with ``fossil'' studies of local massive
elliptical galaxies, which conclude that the stellar populations in such galaxies 
formed early and over a short time period 
(e.g., \citealp{tho05,nel05}).

We present a sample of six $z\sim1.5$ OGs that have both $R\!-\!K^{\prime} > 6$, and 
$J\!-\!K^{\prime}\sim2$.  Finding charts for these OGs are given in Fig.~\ref{fc}.
We have obtained broad-band photometry covering rest-frame UV to visible wavelengths, as well as medium resolution spectroscopy covering the rest-frame UV portion of the spectrum for five of the six sources.  
The spectra provide a more precise measure of the galaxy redshifts than those
determined from the photometry alone, and these redshifts, in turn, allow a more 
accurate determination of the age of the dominant stellar populations.  The spectra are also more sensitive to age diagnostics, such as the Mg {\sc ii} doublet at 2800~{\AA}, as well as to star formation indicators, such as [O {\sc ii}], which are indistinguishable from the underlying continuum in the broad-band photometry.
We examine the stellar populations of these OGs by comparing our observations with the most recent population synthesis models, and place them in context with current semi-analytic models of galaxy formation.  Throughout this paper we assume a flat cosmology with $H_0 = 71$ km s$^{-1}$ Mpc$^{-1}$, $\Omega_M =0.27$ and $\Omega_{\Lambda}=0.73$.

\section{Observations}
Our sample of six OGs at $z\sim1.5$ were found in the fields of radio-loud quasars.
Radio quasar fields were selected based on both the knowledge that radio sources tend to pinpoint more overdense regions of space than their optically selected radio-quiet counterparts 
(e.g., \citealp{bes00,cro01,bes03,bar03,rot05}) 
and the assumption that star formation would proceed most rapidly in the most overdense regions of space, therefore enabling the detection of evolved galaxies at high redshift.  
Details of the photometric selection criteria can be found in 
\citet{sto01} and \citet{sto06}; 
however, we will provide a brief summary here.  

Fields were imaged in the $K^{\prime}$-band, and we looked for objects with magnitudes in the range 
$17 \leq K^{\prime} \leq 19$ 
expected for a galaxy at $z\sim1.5$ that would passively evolve into a $\sim 1$--$5 L^{\ast}$ 
galaxy by the present day.
$J$-band photometry was then obtained to search for all objects with $J\!-\!K^{\prime} \sim 2$, expected for old stellar populations ($> 1$ Gyr) in this redshift range.  Finally $R$-band imaging was acquired to constrain the blueward side of the 4000~{\AA}-break and look for objects with $R\!-\!K^{\prime}>6$.  
\citet{sto06} 
show that these combined $R\!-\!K^{\prime}>6$ and $J\!-\!K^{\prime} \sim 2$ criteria are sufficient to discriminate between heavily dust-reddened star-forming galaxies and passively evolving old stellar populations at $z\sim1.5$ (cf., their Figure 1).  Once these criteria were met, we then obtained additional imaging at blue and intermediate wavelengths and deeper imaging in the near-IR in order to reduce photometric errors and better discriminate between model stellar populations.  

This selection method differs in two important ways from that of 
\citet{dad04} 
who use the $BzK$ criterion to find massive galaxies in the Hubble deep fields.  First, selection around radio sources allows us to select specific redshift ranges where the diagnostics from broad-band filters gives the cleanest separation between dusty starbursts and old stellar populations.  Specifically, for $z\sim1.5$, the 4000~{\AA}-break falls between $I$ and $J$ filters.  Second, by selecting sources in the infrared and using the ERO $R\!-\!K^{\prime}>6$ criterion (as opposed to requiring a $B$-band detection) we are able to reach the faint limiting 
magnitudes required to avoid bias against galaxies with little or no blue flux,
thereby enabling us to pick out the oldest stellar populations more efficiently.  
This method also differs from the distant red galaxy (DRG) selection criterion of $J-K>2.3$ 
\citep{fra03, van03}, 
which is primarily used to select galaxies at $z>2$ without regard for the age of the dominant stellar population.

\subsection{Optical and Near-IR Ground-based Imaging}

Near-IR observations at $J$ and $K^{\prime}$ were obtained on the NASA Infrared Telescope Facility (IRTF) with NSFCam \citep{shu94} for all of the fields.
Deeper follow-up $J$ and $K^{\prime}$ photometry were obtained on Subaru with CISCO \citep{mot02} for TXS\,0145+386 ER1 and ER2, as well as 4C\,15.55 ER2.  Deep $J$-band photometry with CISCO was also obtained for TXS\,1812+412 ER2.
All 6 OGs were observed in the $R$-band using the $2048 \times 2048$  
Tektronix (Tek2048) CCD on the UH 2.2m telescope, the Low-Resolution Imaging Spectrograph (LRIS; \citealt{oke95}) on Keck I or Keck II, and/or the Echellette Spectrograph and Imager (ESI; \citealt{she02}) on Keck II.  
$I$-band imaging was also obtained with the Tek2048 CCD for the fields of TXS\,1812+412 and TXS\,0145+386.  In addition, a deep $z$-band image of the field of TXS\,0145+386 was obtained with LRIS and an RG850 filter.  The combined RG850 filter with the LRIS CCD response is very similar to the standard SDSS $z$ filter profile. 
Deep imaging was obtained for four of the sources (in the fields of B2\,1018+34, TXS\,1211+334, 4C\,15.55, and TXS\,1812+412) in both $g$- and $R_s$ filters on LRIS.  The $R_s$ filter (also referred to as ``script $R$'') is discussed in detail by 
\citet{ste93}; the $R_s$ magnitude can be converted to a standard Cousins $R$ using the $R-I$ color index.  
Table \ref{tab_obs} summarizes all of the observations.  

Flux calibration was carried out using Landolt standard star fields \citep{lan92} for the optical imaging  and UKIRT faint standards \citep{haw01} for the near-IR imaging.  The $z$-band imaging was an exception:  because adequate $z$-band standards were not available at the time the observations were taken, spectrophotometric standards from 
the list of \citet{mas90} were observed instead.  The magnitude zero point (Vega system) was determined by convolving the Vega model SED with the normalized filter + CCD response function.

The data were reduced using standard procedures in IRAF and photometry was performed on the resulting images using apertures that were approximately 4$^{\prime\prime}$ in diameter, the exact aperture size varying from source to source depending on the extent of each individual galaxy.  
Aperture corrections were estimated from isolated stars in the field in order to match the 10$^{\prime\prime}$ apertures that were used on the standard stars.  We used values from 
\citet{sch98} 
to correct for Galactic extinction.  Random errors were determined from the background variance and an additional 5\% 
error was added in quadrature in order to conservatively account for absolute photometric calibration and systematic errors between the ground and space-based observations.  
All photometry is given in the Vega magnitude system.

\begin{deluxetable*}{llcccc}
\tablecolumns{6}
\tablewidth{0pc}
\tablecaption{Summary of Observations}
\tablehead{\colhead{Field\tablenotemark{a}} & \colhead{Date} & \colhead{Telescope} & \colhead{Instrument} & \colhead{Filter} & \colhead{Exposure}\\
\colhead{} & \colhead{(UT)} & \colhead{} & \colhead{} & \colhead{} & \colhead{(s)} }
\startdata
TXS\,0145+386  &   1999 Oct 2  &  Keck II   &   LRIS        &  $R_c$  &  \phn6300 \\
                           &   1999 Oct 12  &  UH 2.2   &   Tek2048  &  $I_c$   &  20400 \\
                           &   1999 Oct 14  &  UH 2.2   &   Tek2048  &  $I_c$   &  20400 \\
                           &   1999 Oct 2 &  Keck II    &   LRIS        &  $z'$     &  13500 \\
                           &   2005 Aug 16 &  Subaru   &   CISCO     &  $J$      &  \phn\phn960 \\
                           &   2005 Aug 16 &  Subaru   &   CISCO     &  $K'$     &  \phn\phn576 \\
                           &   2005 Jan 9 &  HST        &   ACS         &  $F814W$  &  \phn5288 \\
\phantom{X\,0145+38}ER1  &   2005 Jul 16  &  HST        &   NICMOS  &  $F160W$  &  \phn2687 \\
\phantom{X\,0145+38}ER2  &   2005 Jul 16  &  HST        &   NICMOS  &  $F160W$  &  \phn2880 \\
\phantom{X\,0145+38}ER1  &   1999 Nov 3  &  Keck I        &   LRIS  &  Spec  &  16800 \\
\phantom{X\,0145+38}ER1  &   2001 Aug 24  &  Keck I       &   LRIS  &  Spec  &  10800\vspace{2mm}\\
B2\,1018+34    &  2005 Apr 9   &  Keck I      &   LRIS         &  $g$       & \phn5100 \\   
                          &   2001 Apr 29 &  UH 2.2    &   Tek2048  &  $R_c$  &  \phn1800 \\
                           &   2005 Apr 9 &  Keck I     &    LRIS        &  $R_s$  &  \phn5100 \\
                           &   2001 Feb 22 & IRTF       &    NSFCam &  $J$       &  \phn1080 \\
                           &   2001 Feb 21 & IRTF       &    NSFCam &  $K'$      &  \phn\phn540 \\
                           &   2005 Jun 8 &  HST        &   ACS         &  $F814W$  &  \phn5234 \\
\phantom{X\,0145+38}ER2  &   2005 Jun 8  &  HST        &   NICMOS  &  $F160W$  &  \phn2688 \\
\phantom{X\,0145+38}ER2  &   2005 May 3--4  &  Keck II        &   DEIMOS  &  Spec  &  12960\vspace{2mm}\\
TXS\,1211+334  &  2005 Apr 9   &  Keck I      &   LRIS         &  $g$       & \phn5440 \\   
                           &   2002 May 18 & UH 2.2    &    Tek2048 &  $R_c$  &  \phn5400 \\
                           &   2005 Apr 9  & Keck I     &    LRIS        & $R_s$   &  \phn5100 \\
                           &   1998 Feb 13 & IRTF       &    NSFCam & $J$        &  \phn1080 \\
                           &   1998 Feb 14 & IRTF       &    NSFCam & $J$        &  \phn1080 \\
                           &   1998 Feb 12 & IRTF       &    NSFCam & $K'$       &  \phn\phn540 \\
                           &   1998 Feb 14 & IRTF       &    NSFCam & $K'$       &  \phn\phn540 \\
                           &   2005 Dec 4 &  HST        &   ACS         &  $F814W$  &  \phn5234 \\
\phantom{X\,0145+38}ER1  &   2005 Jul 2  &  HST        &   NICMOS  &  $F160W$  &  \phn2688 \\
\phantom{X\,0145+38}ER1  &   2005 May 3--4  &  Keck II        &   DEIMOS  &  Spec  &  11100\vspace{2mm}\\
4C\,15.55          &  2005 Apr 9   &  Keck I      &   LRIS         &  $g$       & \phn5780 \\  
                           &   2002 Aug 7  & Keck II     &   ESI           & $R$       &  \phn2520 \\
                           &   2005 Apr 9  & Keck I     &    LRIS        & $R_s$    &  \phn5700 \\
                           &   1999 Apr 8 & IRTF       &    NSFCam  & $J$        &  \phn3780 \\
                           &   2002 May 31 & Subaru  &    CISCO      & $J$        &  \phn\phn960 \\
                           &   1999 Apr 8  & IRTF      &    NSFCam  & $K'$       &  \phn2160 \\
                           &   2002 May 31 & Subaru  &    CISCO      & $K'$       &  \phn1920 \\
                           &   2005 Feb 28 &  HST        &   ACS         &  $F814W$  &  \phn5000 \\
\phantom{X\,0145+38}ER2  &   2005 Mar 3  &  HST        &   NICMOS  &  $F160W$  &  \phn2688 \\
\phantom{X\,0145+38}ER2  &   2000 Jun 5  &  Keck II        &   ESI  &  Spec  &  18000\vspace{2mm}\\
TXS\,1812+412  &  2005 Apr 9   &  Keck I      &   LRIS         &  $g$       & \phn5440 \\ 
                           &   2002 May 18 & UH 2.2   &    Tek2048   & $R_c$   &  \phn1200 \\
                           &   2005 Apr 9  & Keck I    &    LRIS         & $R_s$   &  \phn5100 \\
                           &   2001 Apr 29  & UH 2.2   &    Tek2048  & $I_c$     &  \phn3240 \\
                           &   2001 Apr 12  & IRTF      &    NSFCam  & $J$        &  \phn1620 \\
                           &   2005 Aug 16 & Subaru   &    CISCO     & $J$        &  \phn1920 \\
                           &   2001 Apr 12 &  IRTF      &    NSFCam  & $K'$       &  \phn1620 \\
                           &   2005 Jan 7 &  HST        &   ACS         &  $F814W$  &  \phn5370 \\
\phantom{X\,0145+38}ER2  &   2005 Jun 8  &  HST        &   NICMOS  &  $F160W$  &  \phn2688 \\
\phantom{X\,0145+38}ER2  &   2005 May 3--4  &  Keck II        &   DEIMOS  &  Spec  &  12560 \\
\enddata
\label{tab_obs}
\tablenotetext{a}{The 
imaging fields were generally centered on the quasar, except for
the {\it HST} NICMOS observations.  Because of the small field of 19\arcsec,
these were centered on the OGs themselves, as indicated in column 1.  For the {\it HST} ACS imaging,
the quasars were placed near the center of the WFC1 CCD.
The slit spectra were centered on the OGs.}
\end{deluxetable*}

\subsection{HST ACS and NICMOS imaging}

We also obtained HST imaging for all six sources.  Details of the data reduction will be described in a future paper 
\citep{mcg07}. 
Photometry was performed on the ACS F814W images using aperture radii that ranged between 1\farcs25 and 1\farcs5.  
These values were chosen to include as much light from the galaxies as possible, while minimizing errors due to background noise.  Preliminary modeling of the galaxy light profiles confirms that $\sim$90\% or more of the total galaxy light falls within the measurement radii; any light falling outside the aperture is within the photometric errors.  
These aperture magnitudes were corrected to a 5$^{\prime\prime}$ radius ``infinite'' aperture 
using correction factors estimated from isolated stars in the ACS field.  The corrections were found to be in good agreement with the values given by 
\citet{sir05} 
based on PSF modeling.  Finally, the aperture corrected instrumental magnitudes were converted to Vega magnitudes using the zero point for the F814W filter given by 
\citet{sir05}.

\begin{figure*}[htb]
\begin{center}
\plotone{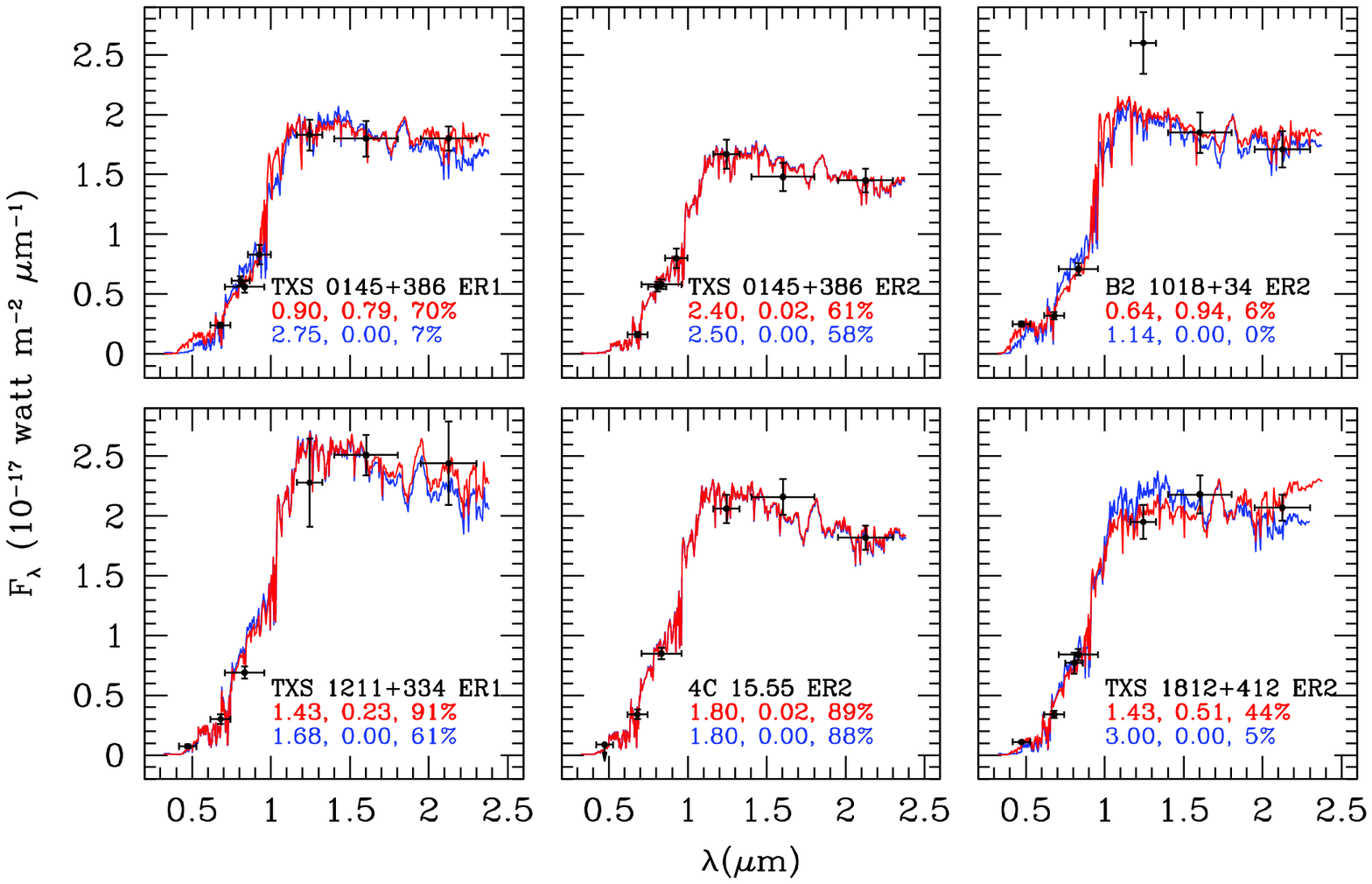}
\end{center}
\caption{Broad-band photometry overplotted with solar metallicity instantaneous burst models from 
\citet{cha07}.  
The best fit models both with (red) and without dust (blue) are shown.  The values listed in the figure correspond to age in Gyr, magnitudes of rest-frame visual extinction, and $P(\chi^2)$, the goodness of the fit.  The model parameters are also given in Table \ref{tab_prop}.\label{sed_fig}}  
\end{figure*}

NICMOS F160W photometry was performed in much the same way as that for the ACS images, however aperture corrections were estimated from TinyTim 
\citep{kri04} 
generated PSFs to determine what percentage of light from the PSF fell outside our measurement aperture.  Galactic extinction corrections were applied to the HST data, and errors were determined based on the rms background variation.  A correction factor was applied to the rms errors for correlated noise due to the drizzling process 
\citep{fru02}, 
and a 5\% absolute photometry error was added in quadrature to the random errors.
Photometry from both ground and space-based observations are given for the six sources in Table \ref{tab_phot}.  The observed broad-band spectral energy distributions (SEDs) along with the best fit models are shown in Figure \ref{sed_fig} and are discussed in detail in \S\ref{sec_photz}.

\begin{deluxetable*}{rcccccccc}
\tabletypesize{\scriptsize}
\tablecolumns{9}
\tablewidth{0pc}
\tablecaption{Photometry of Old Galaxies\tablenotemark{$\dagger$}}
\tablehead{
\colhead{Galaxy} & \colhead{g} & \colhead{R} & \colhead{I} & \colhead{F814W} & \colhead{z}&
\colhead{J}& \colhead{F160W} & \colhead{K$^{\prime}$}
}
\startdata
TXS\,0145+386 ER1 & \nodata &  25.08$\pm$0.10  &  23.28$\pm$0.08  &  23.26$\pm$0.09  &  22.34$\pm$0.11  &  20.54$\pm$0.08  &  19.62$\pm$0.09  &  18.51$\pm$0.06  \\
ER2                             & \nodata  &  25.52$\pm$0.14  &  23.38$\pm$0.08  &  23.21$\pm$0.08  &  22.38$\pm$0.11  &  20.64$\pm$0.08 &  19.83$\pm$0.09  &  18.75$\pm$0.08  \\
B2\,1018+34 ER2      &   25.81$\pm$0.11 &  24.77$\pm$0.09  &  \nodata               &  23.00$\pm$0.08  &  \nodata                &  20.16$\pm$0.11  &  19.59$\pm$0.10  &  18.57$\pm$0.10  \\
TXS\,1211+334 ER1  & 27.14$\pm$0.22 &  24.84$\pm$0.09  &  \nodata               &  23.02$\pm$0.08  & \nodata                &  20.30$\pm$0.18  &  19.26$\pm$0.07  &  18.18$\pm$0.16  \\
4C\,15.55 ER2            & $>26.96$ (3$\sigma$) &  24.69$\pm$0.11  &  \nodata               &  22.80$\pm$0.06  &  \nodata                &  20.41$\pm$0.06  &  19.42$\pm$0.08  &  18.50$\pm$0.06  \\
TXS\,1812+412 ER2 & 26.74$\pm$0.25 &  24.70$\pm$0.09  &  23.04$\pm$0.13  &  22.81$\pm$0.06  &  \nodata                &  20.47$\pm$0.08  &  19.41$\pm$0.08  &  18.36$\pm$0.06  \\

\enddata
\label{tab_phot}
\tablenotetext{$\dagger$}{Values are given in the Vega magnitude system and are corrected for Galactic extinction.}
\end{deluxetable*}

\subsection{Spectra of Five Old Galaxies}
\begin{figure*}[htb]
\begin{center}
\plotone{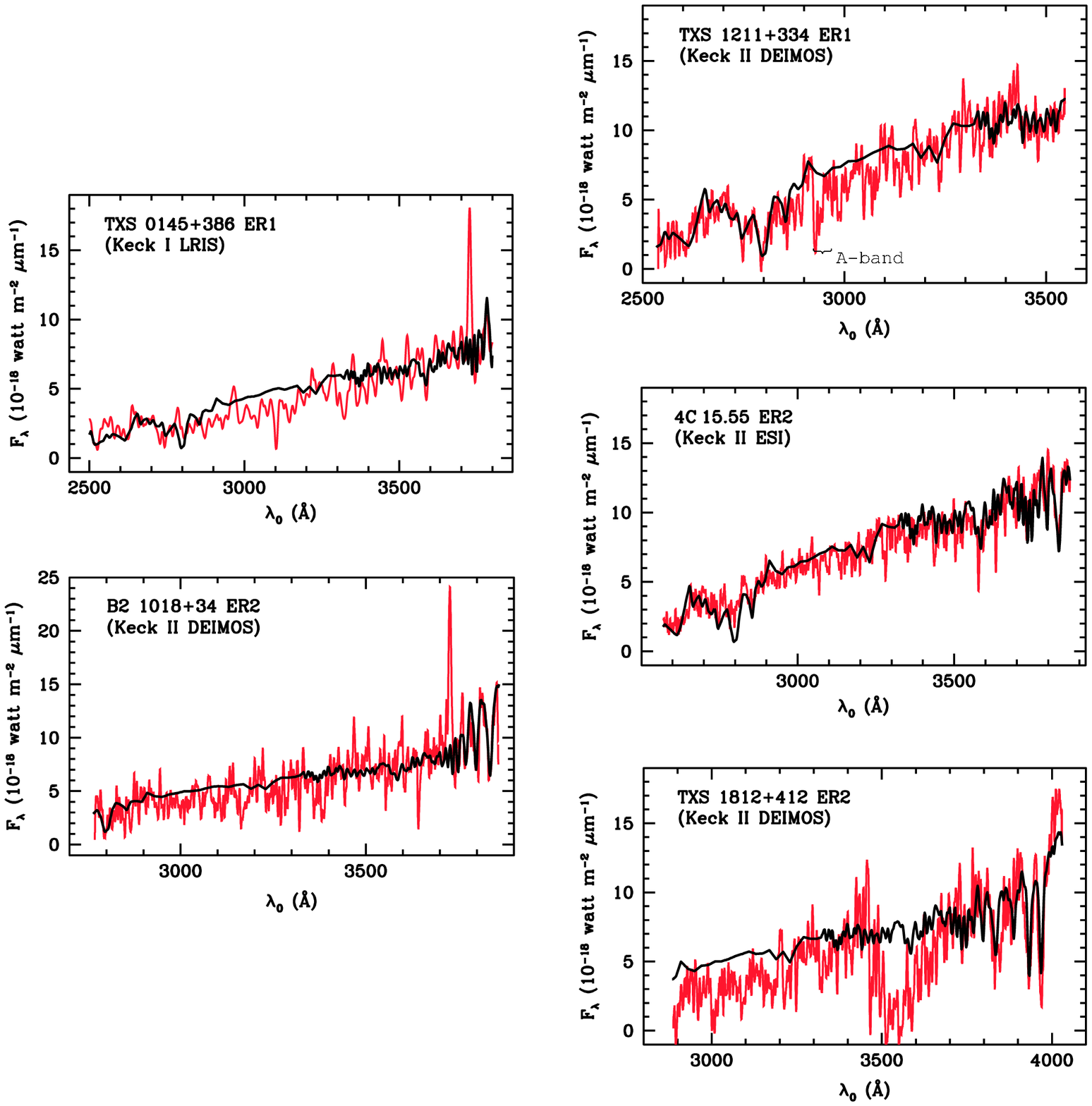}
\caption{The rest-frame UV spectra plotted in red for five of the six galaxies in our sample.  Overplotted in black is the best-fit dust-reddened 
\protect\citet{cha07} 
solar metallicity, instantaneous burst model determined from the photometric SED fitting.  The resolution of the model at $\lambda < 3300$~{\AA} is $\sim20$~{\AA}, while at $\lambda>3300$~{\AA}, it has been smoothed to $\sim8.5$~{\AA} in order to match the observed data.
On the left, TXS\,0145+386 ER1 and B2\,1018+34 ER2 show [O {\sc ii}] emission, which could be due to either recent star formation or active nuclei.  On the right we show the three oldest galaxies.  The Mg {\sc ii} 2800~{\AA} feature is clearly visible in TXS\,1211+334 ER1 and 4C\,15.55 ER2, while Ca {\sc ii} H and K  are visible in TXS\,1812+412 ER2.  Residual A-band atmospheric absorption in the spectrum of TXS\,1211+334 ER1 is marked for reference.  
The $\sim$300~{\AA}-wide region of the spectrum of TXS 1812+412 ER2 around rest-frame 
3550~{\AA} is an artifact.
Redshifts determined from absorption and/ or emission features in each galaxy's spectrum are given in Table \protect\ref{tab_prop}. \label{spectra_fig}}
\end{center}
\end{figure*}

Medium resolution spectra were obtained for five of the six old galaxies in our sample, with either LRIS on Keck I, or ESI or DEIMOS \citep{fab03} on Keck II.  
The LRIS spectrum of TXS\,0145+386 ER1 and the ESI spectrum of 4C\,15.55 ER2 are discussed in detail in 
\citet{sto06}.  
Spectra for B2\,1018+34 ER2, TXS\,1211+334 ER1, and TXS\,1812+412 ER2 were obtained with DEIMOS on May 3-4, 2005.  We used the gold-coated 1200 lines mm$^{-1}$ grating with a central wavelength of $\sim$8000~{\AA} and a slit width of 1\arcsec, which provided a spectral resolution of $\sim$1.1~{\AA}.  We used slits that were tilted by 5 degrees in order to improve sky subtraction.  
To minimize the effects of bad pixels and other chip defects, we designed the slit masks so that we could dither along the slit and maintain alignment.  This was accomplished through the use of small 1\arcsec~alignment boxes at 5 different dither positions 5\arcsec~apart along the slit position angle and corresponding to the position of a faint star near the primary target.  We used this faint star as a quick alignment check for each dither position.  The data were reduced using the DEEP2 pipeline\footnote{The spec2d analysis pipeline was developed at UC Berkeley with support from NSF grant AST-0071048 and can be obtained from http://astro.berkeley.edu/$\sim$cooper/deep/spec2d/.} and the resulting 1-d spectra for each individual dither position were then co-added using the IRAF task \emph{scombine}.  We cleaned and smoothed the spectra to a resolution of $\sim$8.5~{\AA} in order to more easily detect spectral features.  We also obtained slitless spectroscopy of Feige 34 at the position of each primary target in order to flux calibrate the spectra, while the standard Mauna Kea models of atmospheric extinction 
\citep{kri87} 
were used to correct for small ($\pm0.3$) airmass differences.  Finally, the flux calibrated spectra were scaled by small amounts to account for slit losses and to match the $R$- and $I$-band photometry.
The final spectra are shown in Figure \ref{spectra_fig} along with the best fit 
preliminary stellar population models of 
\citet{cha07} \footnote{These models are preliminary in the
sense that they do not yet include the updated Padova 2007 stellar
evolutionary tracks, although they do include the \citet{mar07} 
prescription for thermally-pulsating asymptotic giant branch evolution of
low- and intermediate-mass stars (S. Charlot, private communication).}.

For all five sources, we were able to determine redshifts from either absorption or emission lines in the spectra.  In four out of five cases, these redshifts are close to that of the radio QSO ($\Delta z<0.01$).  For TXS\,1812+412 ER2, the 
redshift is significantly different from that of the radio source. The galaxy therefore cannot be associated with the quasar environment and is, rather, a chance projection.  Spectroscopic redshifts are given along with the QSO redshifts in Table~\ref{tab_prop}.

\begin{deluxetable*}{rccccccc}
\tablecolumns{8}
\tablewidth{0pc}
\tablecaption{Redshifts and Model Parameters}
\tablehead{
\colhead{Galaxy} & \colhead{$z_{QSO}$} & \colhead{$z_{spec}$} & \colhead{Model} & \colhead{Age$-$no dust} & \colhead{Age$-$with dust} & \colhead{A$_V$} & \colhead{Mass} \\
\colhead{} & \colhead{} & \colhead{} & \colhead{} & \colhead{(Gyr)} & \colhead{(Gyr)} & \colhead{(mag)} & \colhead{(10$^{11} M_{\sun}$)}
}
\startdata
TXS\,0145+386 ER1 & 1.442 & 1.4533 & CB07 & 2.75 & 0.90 & 0.79 & 1.74 \\
                                     &             &              & BC03  & 3.00 & 1.28 & 0.57 & 2.00 \\
ER2 & 1.442 & 1.459\tablenotemark{$\ast$} & CB07 & 2.50 & 2.40 & 0.02 & 2.02 \\
         &            &                & BC03 & 2.75 & 2.50 & 0.08 & 2.27 \\
B2\,1018+34 ER2 & 1.4058 & 1.4057 & CB07 &1.14 & 0.64 & 0.94 & 1.42 \\
                                &               &               & BC03 &1.14 & 0.57 & 1.17 & 1.68 \\
TXS\,1211+334 ER1 & 1.5960 & 1.598 & CB07 & 1.68 & 1.43 & 0.23 & 2.90 \\
                                      &              &            & BC03 &  1.80 & 1.43 & 0.29 & 3.15 \\
4C\,15.55 ER2 & 1.406 & 1.412 & CB07 & 1.80 & 1.80 & 0.02 & 1.75 \\
                           &            &             & BC03 & 2.10 & 1.90 & 0.09 & 2.06 \\
TXS\,1812+412 ER2 & 1.564 & 1.290 & CB07 & 3.00 & 1.43 & 0.51 & 1.56 \\
                                     &             &            & BC03 & 3.50 & 1.28 & 0.73 & 1.79 \\
\enddata
\label{tab_prop}
\tablenotetext{$\ast$}{The best-fit photometric redshift is given as no spectroscopy was obtained for this source.}
\end{deluxetable*}

\section{Galaxy Spectral Energy Distributions and Ages}
\label{sec_photz}

For our analysis, we have used 
\citet[CB07]{cha07}
instantaneous burst models with a 
\citet{cha03}
IMF, solar metallicity, and the empirical Calzetti law 
\citep{cal00} 
for dust extinction.  Broad-band photometric fits to the model SEDs were computed with a 
version of the \emph{Hyperz} code 
\citep{bol00} 
that was modified in order to allow finer age steps.  
The fits for all six of the OGs are shown in Figure \ref{sed_fig}.
For all cases except TXS\,0145+386 ER2, the redshift has been fixed to the spectroscopically determined value so that age and reddening are the only free parameters in the fit.    For TXS\,0145+386 ER2, the redshift range was allowed to vary around the values for the QSO and that of ER1, and the best-fit value was found to be $z=1.459$.  

It is important to note that there is quite a bit of uncertainty in the stellar population models themselves, as well as the treatment of dust for high redshift sources 
\citep{pie05}.  
In addition, thermally pulsating asymptotic giant branch (TP-AGB) stars are known to contribute a significant amount of flux in the infrared for stellar populations of intermediate age 
($\sim$1 Gyr; \citealp{mar05}).  
The amount of this contribution is not well known; however, the net effect of including TP-AGB stars is that the population synthesis models become redder overall, allowing younger, less dusty fits to observed galaxy SEDs.  The CB07 models include a larger contribution from TP-AGB stars than any other models to date, including the 
\citeauthor{mar05} \citeyearpar[M05]{mar05}
models, resulting in younger age determinations \citep{bru07}.  In practice, however (as discussed in \S \ref{discussion_sec}), 
we find that by and large when dust is included, the various models all tend to converge towards similar ages.
We therefore believe that by using these improved models, our derived galaxy ages and formation redshifts should be fairly 
reliable estimates.

Another source of uncertainty comes from the well-known age-metallicity degeneracy 
(e.g., \citealp{wor94}).
On average, a factor of 2.5 increase in the metallicity of the model stellar population results in a best-fit age that is  30-50\% younger.  
An important caveat to this simplistic view of the age-metallicity degeneracy is that increasing metallicity can only be balanced by younger ages to a certain point.  Ages much younger than $\sim$1 Gyr do not provide good fits to the data regardless of metallicity as they simply cannot reproduce the sharp inflection at (rest-frame) 4000~{\AA} required by the observations.  
Furthermore, the ages are determined to a large extent by 
both the 4000 \AA\
break and the spectral slope in the rest-frame near-UV.  
For realistic chemical evolution models, 
this spectral region is likely to be dominated
by sub-solar-metallicity stars even if the average stellar metallicity is 
super-solar \citep{mol97, bic04}.  
Briefly, this result follows from
the spread in metallicities that will always be present in the stellar
populations in a galaxy and the lower line blanketing in the
near-UV for lower-metallicity stars.

From Figure \ref{sed_fig} and Table \ref{tab_prop}, it is clear that the majority of these $z\sim1.5$ galaxies are dominated by old stellar populations, implying that star formation ceased at even higher redshifts.  SED fits that do not include 
dust yield older ages; however, in many instances our current photometry is insufficient to distinguish between the best fit dust-reddened and no-dust models. The difference between these models often amounts to only a few percent change in the overall quality of the fit, such that the no-dust model cannot be ruled out with any certainty.  It is also important to note that, while the fits may be marginally better including dust, to some extent this is expected solely because there are more free parameters in the fit.  Longer wavelength data from Spitzer 
(e.g., \citealp{stm07}) 
is needed to clarify how much dust is present in these galaxies and to better constrain their ages.  

For the most part, however, the SEDs paint a consistent picture.  With the exception of B2\,1018+34 ER2, the galaxies are dominated by stellar ages that are $\sim$1 Gyr or greater.  The most extreme case is TXS\,0145+386 ER2, which is dominated by a $\sim$2.5 Gyr stellar population, with little or no dust.  4C\,15.55 ER2 also seems to be entirely free of dust, with an age of 1.8 Gyr.  
Assuming purely passive evolution following a single massive burst of star formation, the implied formation redshifts for these two galaxies are $z_f=3.3$ and $z_f=2.4$, respectively.  More complex and more realistic star formation histories, such as exponentially decaying or truncated star formation, would imply even earlier average formation times.

On the younger edge of the age distribution, it is possible to fit the photometry for both TXS\,0145+386 ER1 and B2\,1018+34 ER2 with models where the break between $I$- and $J$-bands is due to the heavily reddened Balmer break of an intermediate age stellar population instead of a strong 4000~{\AA} break from older stellar populations.
In fact, B2\,1018+34 ER2 was inadvertently allowed a less stringent criterion for $J\!-\!K^{\prime}$ during the initial selection, and we can see clearly that this galaxy does not fit the same mold as the other galaxies.  A significant amount of dust (with a younger underlying stellar population) appears to be almost required in order to fit the bluest ($g$-band) data point, and the $J$-band point is likely contaminated by an emission line due to star formation or an
active nucleus.   H$\beta$ falls within the $J$-band at $z=1.4$; however, the lack of any apparent excess in 
the NICMOS F160W photometry, where we would expect to see H$\alpha$, rules out H$\beta$ emission sufficiently strong to produce the observed excess in the $J$ band.
This excess could, however, be due to strong [O {\sc iii}] emission.  The spectrum of this galaxy (Figure \ref{spectra_fig}) does show [O {\sc ii}] emission, 
and we discuss this, as well as possible [O {\sc iii}] emission 
in more detail in \S\ref{sec_b1018spec}.

A dust reddened model seems favored for TXS\,0145+386 ER1, as well, although the underlying stellar population appears to be older than that of B2\,1018+34 ER2.  A deep blue-band photometry point or deep mid-infrared data would help clarify the amount of dust present, and truly pin down the age of the galaxy.  Over the observed wavelength range, there is very little difference between the dust-reddened and no-dust models.  Given the dust-free model, the implied formation redshift for this source is $z=3.7$.  If we assume, however, that the dust-reddened model is correct (as it exhibits a higher probability than the no-dust model) the implied formation redshift for this source is a modest $z_f=1.9$.  
The spectrum (Figure \ref{spectra_fig}) shows 
that [O {\sc ii}] is present in this galaxy, as well, 
lending credence to the idea that a fair amount of dust could be present.
This [O {\sc ii}] emission was discussed in detail in 
\citet{sto06}, 
and is either due to a weak AGN or a starburst-driven wind since the linewidth implies a velocity that is too great to be due to ordered rotation or internal velocity dispersion alone.

The remaining two sources, TXS\,1211+334 ER1 and TXS\,1812+412 ER2, are well-fit by intermediate amounts of dust (0.23 and 0.51 magnitudes visual absorption, respectively) and underlying old stellar populations.  The difference in $\chi^2$ between the dust-reddened and no dust models,
while not overwhelmingly large, does favor the presence of at least some dust in these galaxies.  In fact, given the presence of [O {\sc ii}] in the previous two galaxies (B2\,1018+34 ER2 and TXS\,0145+386 ER1) and the ability of the population synthesis models to reproduce a moderate amount of dust from the photometric fitting independently of the spectra, gives us confidence that the dust-reddened models provide reasonably good representations of the galaxy SEDs.  It is important to note, however, that models which imply both large inferred ages and high global extinction values would be difficult to explain from a physical standpoint given the lack of evidence for any significant amounts of recent star formation.  In these cases, further observations combined with better modeling would help clarify the situation.
With dust, the implied formation redshifts for TXS\,1211+334 ER1 and TXS\,1812+412 ER2 are $z_f=2.5$ and $z_f=1.9$, respectively.  Without dust, these values shift to $z_f=2.7$ and $z_f=3.4$.

\section{Stellar Populations from Rest-frame UV Spectroscopy}

In the following, we discuss the spectra of the three sources we observed with DEIMOS on Keck II.  For a detailed discussion of the LRIS spectrum of TXS\,0145+386 and the ESI spectrum of 4C\,15.55 ER2, refer to \citet{sto06}.  We have re-fit the spectra for TXS\,0145+386 ER1 and 4C\,15.55 ER2 using the CB07 models determined from the photometry in order to be consistent with the remaining three sources, and these are shown together in Figure \ref{spectra_fig}.

\subsection{{\rm [O {\sc ii}]} emission in B2\,1018+34 ER2}
\label{sec_b1018spec}
The spectrum of B2\,1018+34 ER2 exhibits [O {\sc ii}] emission (3727~{\AA}, Figure \ref{spectra_fig}), pointing to the possibility of recent star formation in 
this galaxy.  The linewidth for the [O {\sc ii}] emission is $\sigma\sim200$ km s$^{-1}$, which is consistent with rotational broadening or internal velocity dispersion.  If we assume the [O {\sc ii}] emission is due to recent star formation, the equivalent width of the [O {\sc ii}] line implies a star formation rate (SFR)
of $\sim 3.5 M_\sun$ yr$^{-1}$, using the relationship given in 
\citet*[eq.~19]{kew04},
which is 
calibrated to agree with $H\alpha$ SFR indicators
and is corrected for extinction by assuming $A_V\sim1$.
Given the best fit SED to the broadband photometry (Figure \ref{sed_fig}), it is likely that 
a significant amount of dust,
up to 1.0 magnitudes rest-frame visual extinction, 
is present in this galaxy,
indicating that this is a reasonable estimate of the star formation rate.  
This SFR is similar to that found for TXS\,0145+386 ER1 \citep{sto06} and to that of dusty, star-forming EROs in the \citet{cim02} sample.
On the other hand, 
if the observed excess in $J$ band is due to [O {\sc iii}] emission, the implied [O {\sc iii}]/[O {\sc ii}] ratio would be $\sim$24, which is too great to be explained by anything other than an AGN.
Such strong [O {\sc iii}] emission would easily be detectible with $J$-band spectroscopy, which
might also have a good chance of detecting broad H$\beta$, if present.

Regardless of whether the [O {\sc ii}] emission is due to star formation or an AGN, it is clear that the underlying stellar population is still relatively old.  A CN absorption feature is visible at 3831~{\AA} as well as a hint of the Mg {\sc ii} line at 2800~{\AA}; however, the latter is very near the edge of the spectral sensitivity.  
Assuming a single stellar population, the implied formation redshift for this source ($z_f \sim 1.7$) is easily within the allowable range of most if not all current SAMs of galaxy formation.  If there is ongoing star formation, it is a relatively small effect by mass, while if there is an AGN present, then it could be the catalyst that quenched the star formation.  The timescale for the latter case appears to be in line with the AGN feedback scenario proposed by 
\citet{cro06}.  

\subsection{Spectrum of TXS\,1211+334 ER1}

The DEIMOS spectrum of TXS\,1211+334 ER1 (Figure \ref{spectra_fig}) shows a very clear absorption trough with a shape characteristic of the Mg {\sc ii} 2800~{\AA} line.  The strength of this absorption line, as well as the slope of the continuum are well fit by a 1.43 Gyr CB07 solar metallicity instantaneous burst model with 0.23 magnitudes of extinction, which was determined through the photometric SED fitting.  A least squares fit to the spectrum alone yields an age of 1.9 Gyr, although the chi-squared has a shallow minimum from 1.5 to 2.0 Gyr.  The redshift derived from the Mg {\sc ii} line is $z=1.598$, which when combined with the age from the photometry 
implies that by $z=2.5$ the galaxy had completely ceased star formation and had accumulated 
nearly $3\times10^{11}M_{\sun}$ (Table \ref{tab_prop}).

\subsection{Spectrum of TXS\,1812+412 ER2}
\label{txs1812_sec}
The spectrum for TXS\,1812+412 ER2 shows a broad absorption feature near 8100~{\AA} (rest-frame $\sim$3550~{\AA}).  
This feature cannot be attributed to atmospheric absorption, both because of the width of the observed feature ($\sim$300~{\AA} observed frame) and because a secondary source on the slit roughly 10\arcsec\ away does not show a similar drop in intensity.  
The feature is also visible in multiple dither positions. However, the broad-band photometry (both $I$-band and F814W observations) do not show any signatures of this absorption.  We therefore are confident that this feature is an instrumental or data-reduction artifact, and we refrain from using this portion of the spectrum in our analysis.

The best-fit CB07 model from the photometry of a 1.43 Gyr stellar population with 0.51 magnitudes of extinction is shown overplotted on the observed spectrum (Figure \ref{spectra_fig}).  Note that while many features, such as Ca {\sc ii} H and K, and CN at 3831~{\AA}, are fairly well reproduced by this model, the depth of the lines and the underlying slope of the continuum are not.  This is suggestive of either an older age or a higher metallicity stellar population.  Recall that we restricted our photometric fits to solar metallicity models.  
A $2.5\times Z_{\odot}$ model with an age of 0.72 Gyr plus 0.76 magnitudes of visual extinction fit to the broad-band photometry provides a slightly better fit to the overall spectrum, as does the dust-free 3.0 Gyr solar metallicity model.  Fitting the spectrum alone, without including longer wavelength photometry, results in a best-fit 2.5$\times Z_{\sun}$ model of 1.4 Gyr, while a $Z_{\sun}$ model yields an age of $\sim$3.5 Gyr.  Given that there are no signatures of 
 young stellar populations or post-starburst features in the spectrum, older ages are slightly favored, however an age of 0.72 Gyr is not 
necessarily precluded by this, especially if 
moderate amounts of dust are present 
as inferred from the best fit photometric model.  
Once again, longer wavelength data is needed in order to better discriminate between models.

\section{Discussion}
\label{discussion_sec}
Previous studies of passive galaxies at $z>1$ have largely relied on the 
\citet[BC03]{bru03} 
stellar population synthesis models, which include a much lower contribution from TP-AGB stars than do the CB07 models, for estimates of galaxy age and mass 
(e.g., \citealp{cim04,mcc04,yan04,dad05,sar05,kri06,sto04,sto06}).  
Others have relied on the M05 models \citep{pie05, mar06}, which are generally more similar to the CB07 models with respect to the TP-AGB evolutionary phase.
In order to compare our results with these studies, we have calculated the best-fit BC03 model SED fits to our photometry,
as well as the best-fit BC03 and M05 models to the galaxy spectra.
The BC03 
model fits to the photometry are presented in Table \ref{tab_prop} along with the best-fit CB07 models.  
When reddening is allowed to be a free parameter, in all but one case (TXS\,0145+386 ER1) the BC03 models require more dust to fit the photometry.  If we hold E(B-V) = 0, the best-fit BC03 models are older than the CB07 no-dust models for five of the six galaxies, but not dramatically so.  
The average increase in age is $\sim$10\% for the BC03 models.  When we include dust, the difference in age between the two models is negligible, and does not show any preferred trend.  This implies that TP-AGB stars do not greatly influence the shape of the SEDs for these galaxies, perhaps because the age is old enough that TP-AGB stars have all evolved to the point where their contribution to the total bolometric luminosity is dwarfed by lower mass, main sequence stars.  In addition, the rest-frame near-IR, where the contribution from TP-AGB stars is greatest, is not explored by our observations.  The clearest difference between the BC03 and CB07 models is that our inferred galaxy masses are slightly smaller using the CB07 models ($\sim$10\%), 
and in general we find that the $\chi^2$ fits are better with the CB07 models than with the BC03 models.

When we constrain our model fitting to the spectra alone, we get results that are typically within a few hundred Myr of the values determined using similar models with the broad-band photometry.  The only exception to this is that for 4C\,15.55 ER2, the best-fit CB07 age from the spectrum is 1.1 Gyr, as opposed to an age of 1.8 Gyr from the photometry.  The 1.1 Gyr model, however, provides an extremely poor fit to the near-IR photometry, with a P($\chi^2$) close to 0\%, while the chi squared fit to the spectrum has a broad minimum from 1.0 to 1.8 Gyr.  The older age from the photometry is therefore preferred, as it is the most consistent with both datasets.  When fitting the spectra, the difference in $\chi^2$ between dust and no-dust models is often very small, as would be expected given the small wavelength range of the fit; the presence of dust effectively just changes the scaling required to fit the data.  This is exemplified in the fit for TXS 0145+386 ER1, where the difference in $\chi^2$ between the no-dust (2.75 Gyr) and dust-reddened (1.1 Gyr) models amounts to only a 7\% difference in the overall quality of the fit.  We find that the M05 models 
agree well with the CB07 models, however the average best-fit age is slightly older when using the M05 models, and the agreement with the near-IR photometry is generally worse.  Furthermore, when we allow metallicity to vary, the best-fit models favor solar-metallicity in all but one case (TXS\,1812+412 ER2, as discussed in \S \ref{txs1812_sec}).

The seven passive galaxies found in the Hubble Ultra Deep Field by 
\citet{dad00,dad05} 
and re-examined by 
\citet{mar06}, 
have similar formation redshifts to five of our six $z=1.5$ galaxies 
and therefore provide a fair comparison to the galaxies in the present study.  Passive galaxies in the \citeauthor{dad00} sample are generally a factor of two less massive than ours and a few exhibit a brighter blue tail or bump in the SED which yield ages as young as 0.3 Gyr.  
Several of their objects at $z<2$, however, have properties similar to the ones in this study, which expands on the number of known high redshift evolved galaxies.
At even higher redshifts ($2<z<2.7$),
\citet{kri06} 
discuss nine massive galaxies whose stellar continua and Balmer breaks are detected in near-IR spectroscopy, but show no trace of emission lines from young star formation.  These galaxies are fit by stellar populations with $\sim$1.0 magnitudes visual extinction that are on average $\sim$500 Myr old, which imply formation redshifts of $z_f\sim$3.  The galaxies in the present study could very well be the
passively evolved descendants of such a population, with the proviso that much of the dust present in the 
\citeauthor{kri06}
sample would need to be expelled rather than consumed in star formation between $2.5<z<1.5$ in order to match both the lower dust content and the continued passive evolution seen in our $z=1.5$ galaxy sample.  As a cautionary note, however, the ages for these $2<z<2.7$ galaxies fall within the range where the contribution from TP-AGB stars may be much more significant than for our $\sim$2 Gyr-old $z=1.5$ galaxies, and therefore care must be taken in comparing the two populations.  
Use of models with improved treatments of TP-AGB stars such as M05 or CB07 could yield lower ages for these sources.  Alternatively, 
the M05 or CB07 models could require less dust, thereby bringing these observations into even better agreement with the idea that they are progenitors to galaxies at $z=1.5$ both in the present study and in the \citet{dad05} sample.  
Finally, galaxies such as the evolved disk at $z=2.5$ found by 
\citet{sto04} 
appear to be altogether a much more extreme case of early and rapid formation of the most massive galaxies, with an implied formation $<1$ Gyr after the Big Bang.  It is unclear how common objects such as these are, and it will be important to continue searching for the oldest stellar populations at higher redshifts in order to further constrain galaxy formation models.

\section{Conclusions}
We have studied a sample of six evolved galaxies at $z\sim1.5$ with broadband photometry covering the rest-frame UV to visible wavelengths, as well as medium resolution spectroscopy from Keck covering the rest-frame UV portion of the spectrum.  Using the most recent stellar population synthesis models of 
\citet{cha07} 
which include an improved treatment of TP-AGB stars, we find ages for these galaxies that imply formation redshifts roughly between $2<z<4$.  Simple SAMs are unable to reproduce massive galaxies at such high redshifts, however, models which employ feedback from AGN may not be as inconsistent with these observations.  
Galaxies such as TXS\,0145+386 ER1 and TXS\,1812+412 ER2, under the assumption of low dust content, test the limit of galaxy formation theories by requiring massive galaxies to be present at $z>3.4$.  
If dustier models are preferred for these galaxies, however, the formation redshifts are more in line with values expected in the AGN feedback scenario.
Clearly, longer wavelength data from Spitzer is needed to pin down the dust content and age of these galaxies.  
Irrespective of dust, the best-fit spectral synthesis model for TXS\,0145+386 ER2 implies a 
$2\times10^{11}M_{\sun}$ galaxy 
that completed star formation by at least
$z=3.3$.  This may be difficult to explain even with AGN feedback.

[O {\sc ii}] is observed in two of the sources (TXS\,0145+386 ER1 and B2\,1018+34 ER2), which may either be an indication of lingering star formation or 
the remains of weak AGN.  If the latter scenario is correct, it may give an indication over what timescale AGN activity is observable after star formation has ceased, which may prove useful for testing the AGN feedback scenario. 

Overall, these observations imply that massive galaxies did in fact exist at very early times in the universe.  Even given the uncertainties in the population synthesis modeling, and the trend toward lower ages for models with improved treatments of TP-AGB stars, we are faced with several examples of massive galaxies ($>10^{11}M_{\sun}$) that appear to have been passively evolving
since $z>2$.  We will explore how these galaxies may have formed in a future paper 
\citep{mcg07} 
through a high-resolution morphological study.

\acknowledgments

We are grateful to S. Charlot and G. Bruzual for providing access to their models prior to publication.  An anonymous referee provided useful suggestions that helped improve the paper.  EJM thanks the UC Davis cosmology group for their hospitality.  This research was partially supported by NSF grant AST 03-07335 and by grant GO-10418 from the Space Telescope Science Institute, which is operated by
the Association of Universities for Research in Astronomy, Inc., under NASA contract
NAS 5-56555.
The authors recognize the very significant cultural role that the summit of Mauna Kea has within the indigenous Hawaiian community and are grateful to have had the opportunity to conduct observations from it.


\begin{thebibliography}{}


\bibitem[Barr et al.(2003)]{bar03}
Barr, J. M., Bremer, M. N., Baker, J. C., \& Lehnert, M. D. 2003, \mnras, 346, 229

\bibitem[Best(2000)]{bes00}
Best, P. N. 2000, \mnras, 317, 720

\bibitem[Best et al.(2003)]{bes03}
Best, P. N., Lehnert, M. D., Miley, G. K., \& R\"ottgering, H. J. A. 2003, \mnras, 343, 1

\bibitem[Bicker et al.(2004)]{bic04} Bicker, J., Fritze-v. Alvensleben, U., M\"{o}ller, C. S., \& Fricke, K. J. 2004, \aap, 413, 37

\bibitem[Bolzonella et al.(2000) Bolzonella, Miralles, \& Pell\'{o}]{bol00}
Bolzonella, M., Miralles, J.-M., \& Pell\'o, R. 2000, \aap, 363, 476

\bibitem[Bower et al.(2006)]{bow06}
Bower, R. G., Benson, A. J., Malbon, R., Helly, J. C., Frenk, C. S., Baugh, C. M., Cole, S., \& Lacey, C. G. 2006, \mnras, 370, 645

\bibitem[Bruzual(2007)]{bru07}
Bruzual, A. G. 2007, in IAU Symp. 241, Stellar Populations as Building Blocks of Galaxies, eds. A. Vazdekis \& R. Peletier (Cambridge: Cambridge University Press), in press [astro-ph/0703052]

\bibitem[Bruzual \& Charlot(2003)]{bru03}
Bruzual, G., \& Charlot, S. 2003, \mnras, 344, 1000

\bibitem[Calzetti et al.(2000)]{cal00}
Calzetti, D., Armus, L., Bohlin, R. C., Kinney, A. L., Koornneef, J., \& Storchi-Bergmann, T. 2000, \apj, 533, 682

\bibitem[Chabrier(2003)]{cha03}
Chabrier, G. 2003, \pasp, 115, 763

\bibitem[Charlot \& Bruzual(2007)]{cha07}
Charlot, S. \& Bruzual, A. G. 2007, \mnras, in preparation

\bibitem[Cimatti et al.(2002)]{cim02}
Cimatti, A., et al. 2002, \aap, 381, L68

\bibitem[Cimatti et al.(2004)]{cim04}
Cimatti, A., et al. 2004, \nat, 430, 184

\bibitem[Cowie et al.(1996)]{cow96}
Cowie, L. L., Songaila, A., Hu, E. M., \& Cohen, J. G. 1996, \aj, 112, 839

\bibitem[Croom et al.(2001)]{cro01}
Croom, S. M., Shanks, T., Boyle, B. J., Smith, R. J., Miller, L., Loaring, N. S., \& Hoyle, F. 2001, \mnras, 325, 483

\bibitem[Croton et al.(2006)]{cro06}
Croton, D. J., et al. 2006, \mnras, 365, 11

\bibitem[Daddi et al.(2000)]{dad00}
Daddi, E., Cimatti, A., Pozzetti, L., Hoekstra, H., R\"ottgering, H. J. A., Renzini, A., Zamorani, G., \& Mannucci, F. 2000, \aap, 361, 535

\bibitem[Daddi et al.(2004)]{dad04}
Daddi, E., Cimatti, A., Renzini, A., Fontana, A., Mignoli, M., Pozzetti, L., Tozzi, P., \& Zamorani, G. 2004, \apj, 617, 746

\bibitem[Daddi et al.(2005)]{dad05}
Daddi, E., et al. 2005, \apj, 626, 680

\bibitem[De Lucia et al.(2006)]{del06}
De Lucia, G., Springel, V., White, S. D. M., Croton, D., \& Kauffmann, G., \mnras, 366, 499

\bibitem[Faber et al.(2003)]{fab03}
Faber, S. M., et al. 2003, Proc. SPIE, 4841, 1657

\bibitem[Franx et al.(2003)]{fra03}
Franx, M., et al. 2003, \apjl, 587, L79

\bibitem[Fruchter \& Hook(2002)]{fru02}
Fruchter, A. S. \& Hook, R. N. 2002, \pasp, 114, 144

\bibitem[Fu et al.(2005) Fu, Stockton, \& Liu]{fu05}
Fu, H., Stockton, A., \& Liu, M. 2005, \apj, 632, 831

\bibitem[Granato et al.(2004)]{gra04}
Granato, G. L., De Zotti, G., Silva, L., Bressan, A., \& Danese, L. 2004, \apj, 600, 580

\bibitem[Hawarden et al.(2001)]{haw01}
Hawarden, T. G., Leggett, S. K., Letawsky, M. B., Ballantyne, D. R., \& Casali, M. M. 2001, \mnras, 325, 563

\bibitem[Iye et al.(2003)]{iye03}
Iye, M., et al.~2003, ApJ, 590, 770

\bibitem[Juneau et al.(2005)]{jun05}
Juneau, S., et al. 2005, \apjl, 619, L135

\bibitem[Kauffmann \& Haehnelt(2000)]{kau00}
Kauffmann, G., \& Haehnelt, M. 2000, \mnras, 311, 576

\bibitem[Kewley et al.(2004) Kewley, Geller, \& Jansen]{kew04}
Kewley, L. J., Geller, M. J., \& Jansen, R. A. 2004, \aj, 127, 2002

\bibitem[Kriek et al.(2006)]{kri06}
Kriek, M., et al. 2006, \apjl, 649, L71

\bibitem[Krisciunas et al.(1987)]{kri87}
Krisciunas, K., et al. 1987, \pasp, 99, 887

\bibitem[Krist(2004)]{kri04}
Krist, J. 2004, {http://www.stsci.edu/software/tinytim/tinytim.html}

\bibitem[Landolt(1992)]{lan92}
Landolt, A. U. 1992, \aj, 104, 340

\bibitem[Liu et al.(2000)]{liu00}
Liu, M. C., Dey, A., Graham, J. R., Bundy, K. A., Steidel, C. C., Adelberger, K., Dickinson, M. E. 2000, \aj, 199, 2556

\bibitem[Maraston(2005)]{mar05}
Maraston, C. 2005, \mnras, 362, 799

\bibitem[Maraston et al.(2006)]{mar06}
Maraston, C., Daddi, E., Renzini, A., Cimatti, A., Dickinson, M., Papovich, C., Pasquali, A., \& Pirzkal, N. 2006, \apj, 652, 85

\bibitem[Marigo \& Girardi(2007)]{mar07} Marigo, P. \& Girardi, L. 2007,
   \aap, in press [astro-ph/0703139]

\bibitem[Massey \& Gronwall(1990)]{mas90} Massey, P., \& Gronwall, C., 
   \apj, 358, 344

\bibitem[McCarthy et al.(2004)]{mcc04}
McCarthy, P. J., et al. 2004, \apjl, 614, L9

\bibitem[McGrath et al.(2007)]{mcg07}
McGrath, E. J., Stockton, A., Canalizo, G., Iye, M., \& Maihara, T. 2007, \apj, submitted

\bibitem[M\"{o}ller, Fritze-v. Alvensleben, \& Fricke(1997)]{mol97} M\"{o}ller, C. S., Fritze-v. Alvensleben, U., \& Fricke, K. J. 1997, \aap, 317, 676 

\bibitem[Motohara et al.(2002)]{mot02}
Motohara, K., et al. 2002, \pasj, 54, 315

\bibitem[Nelan et al.(2005)]{nel05} Nelan, J. E., Smith, R. J., Hudson, M. J.,
   Wegner, G. A., Lucey, J. R., Moore, S. A. W., Quinney, S. J., \& Suntzeff,
   N. B. 2005, \apj, 632, 137

\bibitem[Oke et al.(1995)]{oke95} Oke, J. B., et al. 1995, \pasp, 107, 375

\bibitem[Pierini et al.(2005)]{pie05}
Pierini, D., Maraston, C., Gordon, K. D., \& Witt, A. N. 2005, \mnras, 363, 131

\bibitem[R\"ottgering et al.(2005)]{rot05}
R\"ottgering, H., de Breuck, C., Daddi, E., Kurk, J., Miley, G., Pentericci, L., Overzier, R., \& Venemans, B. 2005, in Multiwavelength Mapping of Galaxy Formation and Evolution, ed. A. Renzini \& R. Bender (Berlin: Springer), 50

\bibitem[Saracco et al.(2005)]{sar05}
Saracco, P., et al. 2005, \mnras, 357, L40

\bibitem[Schlegel et al.(1998) Schlegel, Finkbeiner, \& Davis]{sch98}
Schlegel, D. J., Finkbeiner, D., \& Davis, M. 1998, \apj, 500, 525

\bibitem[Sheinis et al.(2002)]{she02}
Sheinis, A. I., Bolte, M., Epps, H. W., Kibrick, R. I., Miller, J. S., Radovan, M. V., Bigelow, B. C., \& Sutin, B. M. 2002, \pasp, 114, 851

\bibitem[Shure et al.(1994)]{shu94}
Shure, M. A., Toomey, D. W., Rayner, J. T., Onaka, P. M., \& Denault, A. J. 1994, Proc.~SPIE, 2198, 614

\bibitem[Sirianni et al.(2005)]{sir05}
Sirianni, M., et al. 2005, \pasp, 117, 1049

\bibitem[Steidel \& Hamilton(1993)]{ste93}
Steidel, C. C. \& Hamilton, D. 1993, \aj, 105, 2017

\bibitem[Stockton(2001)]{sto01}
Stockton, A. 2001, in ASP Conf. Ser., 245, Astrophysical Ages and Timescales, ed. T. von Hippel, C. Simpson, \& N. Manset (San Francisco: ASP), 517

\bibitem[Stockton et al.(2004) Stockton, Canalizo, \& Maihara]{sto04}
Stockton, A., Canalizo. G., \& Maihara, T. 2004, \apj, 605, 37

\bibitem[Stockton \& McGrath(2007)]{stm07}
Stockton, A., \& McGrath, E. 2007, ASP Conf. Ser., Cosmic Frontiers, in press [astro-ph/0702130]

\bibitem[Stockton et al.(2006) Stockton, McGrath, \& Canalizo]{sto06}
Stockton, A., McGrath, E., \& Canalizo, G. 2006, \apj, 650, 706

\bibitem[Stockton et al.(2007)]{sto07}
Stockton, A., McGrath, E., Canalizo, G., Iye, M., \& Maihara, T. 2007, \apj, submitted

\bibitem[Thomas \& Kauffmann(1999)]{tho99}
Thomas, D., \& Kauffmann, G. 1999, in ASP Conf. Ser. 192, Spectrophotometric Dating of Stars and Galaxies, ed. I. Hubeny, S. R. Heap, \& R. H. Cornett, (San Francisco: ASP), 261

\bibitem[Thomas et al.(2005)]{tho05} Thomas, D., Maraston, C., Bender, R.,
   Mendez de Oliveira, C. 2005, \apj, 621, 673

\bibitem[van Dokkum et al.(2003)]{van03}
van Dokkum, P., et al. 2003, \apjl, 587, L83

\bibitem[Worthey(1994)]{wor94}
Worthey, G. 1994, \apjs, 95, 107

\bibitem[Yan et al.(2004) Yan, Thompson, \& Soifer]{yan04}
Yan, L., Thompson, D., \& Soifer, B. T. 2004, \aj, 127, 1274


\end{thebibliography}
\end{document}